\documentclass{kluwer}
\usepackage{epsf,amssymb}
\usepackage{graphics}
\usepackage{graphicx}
\usepackage{epsfig}

\newdisplay{guess}{Conjecture}

\begin{document}
\begin{article}

  \begin{opening}         

    \title{A thin ring model for the OH megamaser in IIIZw35} 
    \author{Rodrigo \surname{Parra}}
    \author{John \surname{Conway}}
    \institute{Onsala Space Observatory, Sweden}
    \author{Moshe \surname{Elitzur}}
    \institute{Department of Physics and Astronomy, University of Kentucky, USA}
    \runningauthor{Parra, Conway and Elitzur}
    \runningtitle{A thin ring model for the OH megamaser in IIIZw35}
    \date{April 15, 2004}

   \begin{abstract}
   We present a model for the OH megamaser emission in the starburst galaxy
   IIIZw35. The observed diffuse and compact OH maser components in this
   source are explained by a single phase of unsaturated clumpy gas
   distributed in a thin ring structure and amplifying background
   continuum. We emphasize the importance of clumpiness in the OH masing
   medium, an effect that has not been fully appreciated previously. 
   
   The model explains why multiple bright spots are seen only at the ring tangents
   while smoother emission is found elsewhere. Both the observed velocity
   gradients and the line to  continuum ratios around the ring enquire a
   geometry where most of the seed photons come from a continuum emission
   which lies outside the OH ring. To explain both the OH and continuum
   brightness, free-free absorbing gas is required along the ring axis to
   partially absorb the far side of the ring. It is proposed that the required
   geometry arises from an inwardly propagating ring of starburst activity.
    \end{abstract}

  \end{opening}

\section{Introduction}
Extragalactic OH megamasers usually present both extended diffuse emission and
very compact and strong features. One of the clearest cases of an OH megamaser
showing both compact and diffuse masers is \mbox{IIIZw35}. Two groups of compact 
masers were detected in VLBI (Trotter et al. 1998; Diamond et al. 1989)
recovering nearly half of the total emission seen at MERLIN scales (Montgomery
and Cohen 1992). Aiming to determine the location of the missing diffuse
component MERLIN+EVN observations were conducted on this source (Pihlstr\"{o}m
et al. 2001). These observations revealed the best example of a rotating OH
maser ring that has yet been found. 

It has been claimed that the two-phase characteristic of the maser features
observed toward IIIZw35 is due to two coexisting phases of the OH gas (Diamond
et al. 1989), one producing the smooth emission and the other responsible for
the very strong and compact features. According to this hypothesis, and
making the usual assumption that the diffuse component maser is operating in
the unsaturated regime, the Line to Continuum Ratio (LCR) of each phase is
exponentially related to the path length through the amplifying gas.
In the particular case of the smooth component in IIIZw35, the LCR in the
east/west and north/south regions has been observed to be around 10 and 1000
respectively. Additionally, the ratio of the path lengths between these
regions is estimated to be approximately 8 therefore, the ratio of their LCRs
should be in the order of $10^{7}$.
Instead in order to match the observations, a clumpy
medium is proposed so the column density of amplifying gas is proportional to
the number of overlapped clumps along the line of sight. This number is the
result of a random process that yields a mean which is linearly related to the
projected path length through the gas.

\begin{figure}
\begin{center}
$\begin{array}{c@{\hspace{1cm}}c}
\includegraphics[width=0.4\columnwidth]{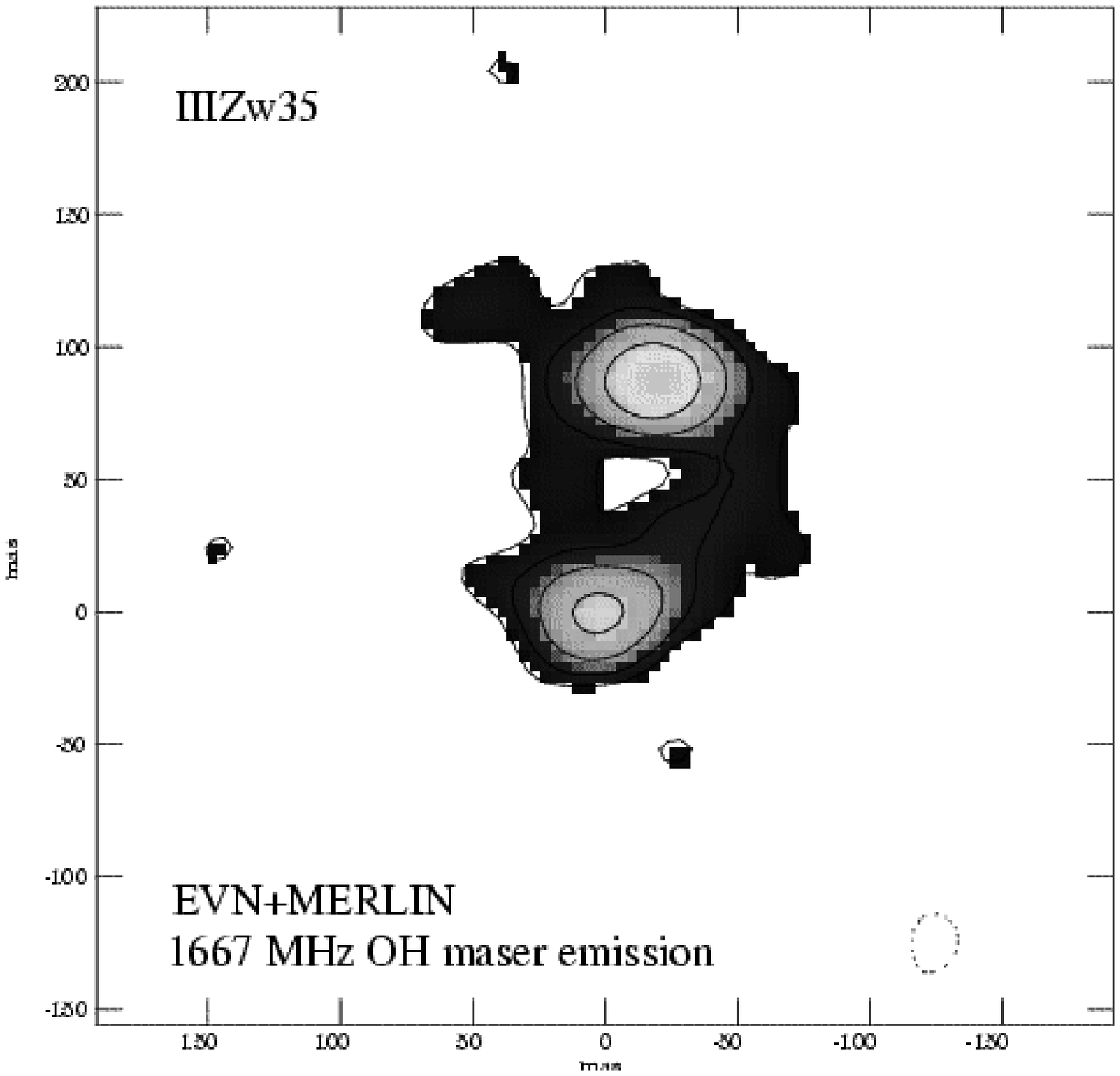}&
\includegraphics[width=0.4\columnwidth]{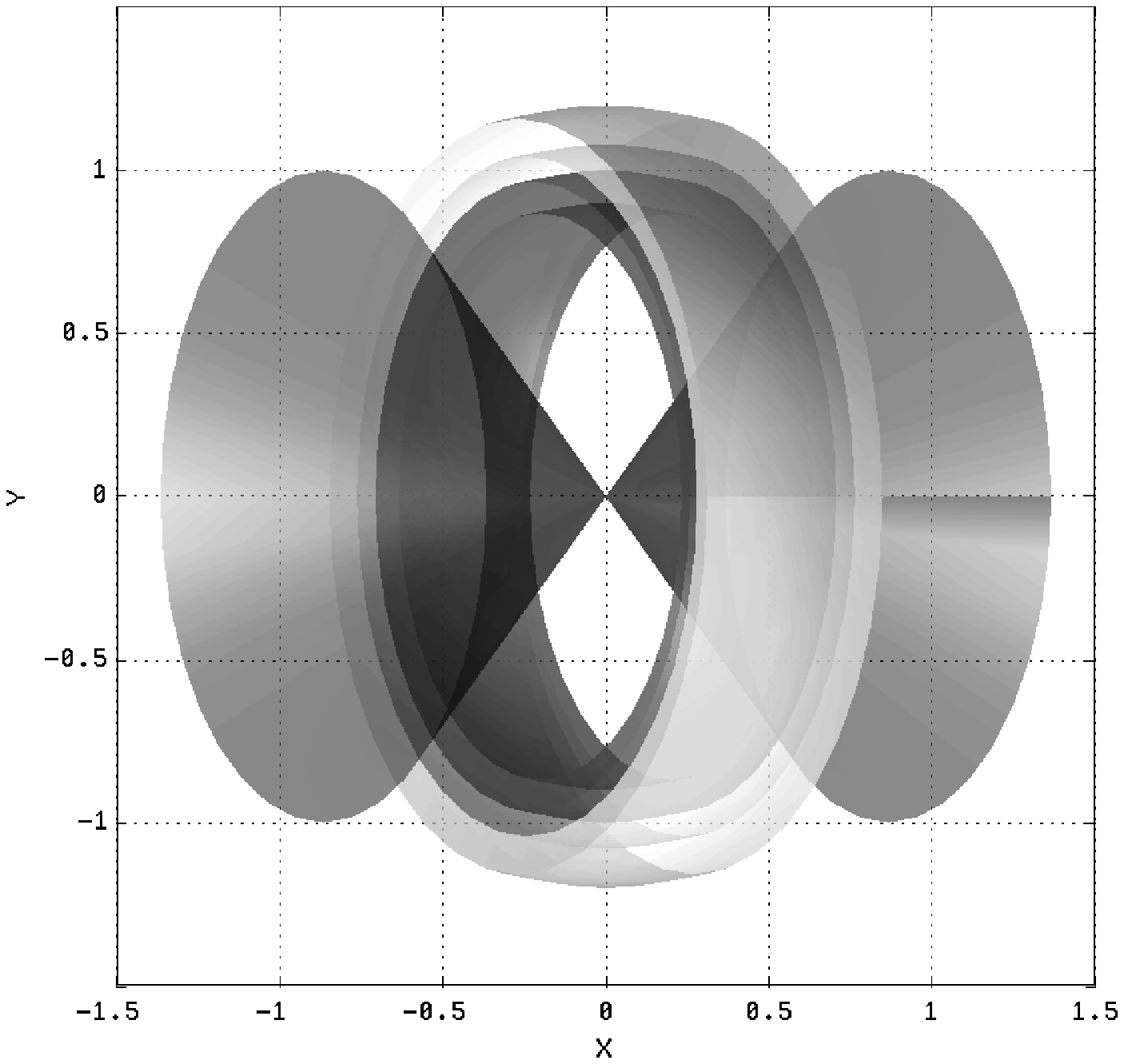}
\end{array}$
\end{center}
\caption{{\bf Left: }Velocity averaged OH emission as seen at EVN+MERLIN
  resolution (Pihstr\"{o}m et al. 2001). The contours are -1, 1, 2, 4 and 8
  times the 3$\sigma$ rms noise of 2.7 mJy/beam. {\bf Right: }The proposed
  geometry. The OH gas is concentrated within the dark gray ring and most of
  the continuum is emitted within the surrounding light gray ring. The medium
  gray cones indicate the region where free-free absorption occurs. The units
  are normalized to the outer OH radius R$_{\mathrm{o}}$.}
\label{fi:figure1}
\end{figure}

\section{Description of the model}
\subsection{Geometry}
We propose that the conditions for the OH gas to produce maser amplification
are satisfied within a thin ring structure with inner and outer radii
R$_{\mathrm{i}}$ and R$_{\mathrm{o}}$. Surrounding the OH, there is another
concentric thin ring of radius R$_{\mathrm{c}}$ where most of the continuum emission is produced. In
addition, free-free absorption cones along the axis are assumed to account for
the east-west asymmetry in the brightness distribution. This situation is
depicted in Figure \ref{fi:figure1} and the observed geometrical parameters
are summarized in Table \ref{ta:table1}. 

\begin{table}
\caption{Observed geometrical parameters.}
\begin{minipage}{10cm}
\begin{tabular}{crl}
\hline
\bf Parameter&\bf Value&\bf Reference\\
\hline
$\Delta$R&3 Pc\phantom{g}&Diamond et al. 1998; Trotter et al. 1989\\
$\Delta$H&6 Pc\phantom{g}&Diamond et al. 1998; Trotter et al. 1989\\
R$_{\mathrm{i}}$&22 Pc\phantom{g}&Trotter et al. 1989\\
R$_{\mathrm{o}}$&25 Pc\phantom{g}&Pihlstr\"{o}m et al. 2001\\
R$_{\mathrm{c}}$&50 Pc\phantom{g}&Pihlstr\"{o}m et al. 2001\\
$\theta$&60 Deg&Pihlstr\"{o}m et al. 2001\\
\hline
\end{tabular}
\end{minipage}
\label{ta:table1}
\end{table}

The fact that the northern and southern compact masers seem to be elongated
toward the east (Diamond et al. 1998; Trotter et al. 1989) is explained in our
model by the larger amount of continuum behind the OH in these
regions. Accordingly, the apparent east-west extention of the feature
corresponds to half of the ring width $\Delta$H whereas the north-south
extention matches the difference between the outer and inner radii
$\Delta$R. The angle $\theta$ between the axis of symmetry of the disk and the
direction of the observer is estimated from the aspect ratio of the projected
ring (Pihlstr\"{o}m et al. 2001).

\subsection{Clumpy medium analysis}
The \emph{clumpiness} of the medium is defined by the parameter $\bar{n}$
which defines the average number of clumps per cloud velocity width found in
the radial direction within the ring. The average number of clumps in the
direction of the observer is given by $\lambda(x,y)=\bar{n}\cdot L(x,y)/\Delta
R$ where $L(x,y)$ is the projected path length through the OH gas.
For any given line of sight, the number of observed spatial overlappings per
velocity range is drawn from a Poisson random process of parameter $\lambda$
with Probability Mass Function $P_{\lambda}(n)$. If all the clumps have the
same optical depth $\tau_{0}$, then the expected amplification $G$ along the
line is given by the sum:

\begin{equation}
G=\sum_{n=0}^{N}P_{\lambda}(n)\cdot e^{n\tau_{0}}
\label{eq:sum}
\end{equation}

\noindent
which in the limit of $N\rightarrow\infty$ converges to
$e^{\lambda(e^{\tau_{0}}-1)}$ (See Conway et al. in this volume). This result
is inappropriate in our context since the maximum number of aligned clumps $N$
is upper bounded either by geometrical or statistical constrains. For these reasons, the mean does
not constitute a good estimator for the observations. A better choice could be
to use the median instead. A simplified approximation is to define an expected
maximum $N$ in order to truncate the sum in (\ref{eq:sum}). Given a set of $M$
independent realizations of the process, corresponding to $M$ independent
lines of sight within the beam, and requiring at least one occurrence
of $N$ overlaps to be observed, the expression for the maximum $N$ becomes:

\begin{equation}
N_{\lambda}(M)=P_{\lambda}^{-1}\left(1-\frac{1}{M}\right)
\label{eq:inverse}
\end{equation}

\noindent
Where $P_{\lambda}^{-1}(x)$ is the Inverse Poisson Cumulative Distribution of
parameter $\lambda$, and $M$ is given by the ratio of the areas of the
EVN+MERLIN to the VLBI synthesized beams when the clumps are assumed to be
just resolved at the latter resolution. For the proposed geometry we define
the following quantities and their observed values:

\begin{equation}
\begin{array}{ccll}
G_{1}&=&\displaystyle \sum_{n=0}^{N_{1}}P_{\lambda_{1}}(n)\cdot
e^{n\tau_{0}}\approx 10&\textrm{\qquad Gain in Region 1}\nonumber\\
G_{2}&=&\displaystyle \sum_{n=0}^{N_{2}}P_{\lambda_{2}}(n)\cdot
e^{n\tau_{0}}\approx 50&\textrm{\qquad Gain in Region 2}\label{eq:parameters}\\
\\
G_{P}&=&\displaystyle e^{N_{2}\tau_{0}}\geq 500&\textrm{\qquad Peak gain in
  Region 2}\nonumber\\
\end{array}
\end{equation}

\noindent
Where $N_{1}$ and $N_{2}$ are the maximum number of cloud overlaps in
regions 1 and 2 given by (\ref{eq:inverse}) and
$\lambda_{2}=K\cdot\lambda_{1}$ with $K$ being the ratio of the averaged path
lengths within both regions. Each of these equations define a
locus in the $\lambda_{1}-\tau_{0}$ plane where the observed
values are matched. The intersection of those loci corresponds to the possible
combinations of $\lambda_{1}$ and $\tau_{0}$ satisfying all three conditions
simultaneously. Solving numerically, the solution is found to be
$\lambda_{1}=0.67$ and $\tau_{0}=1.69$ which corresponds with a mean free path
between clumps $\ell \approx 1.72$ pc.

\begin{figure}[ht!]
\begin{center}
$\begin{array}{c@{\hspace{1cm}}c}
\includegraphics[angle=0,width=0.30\columnwidth]{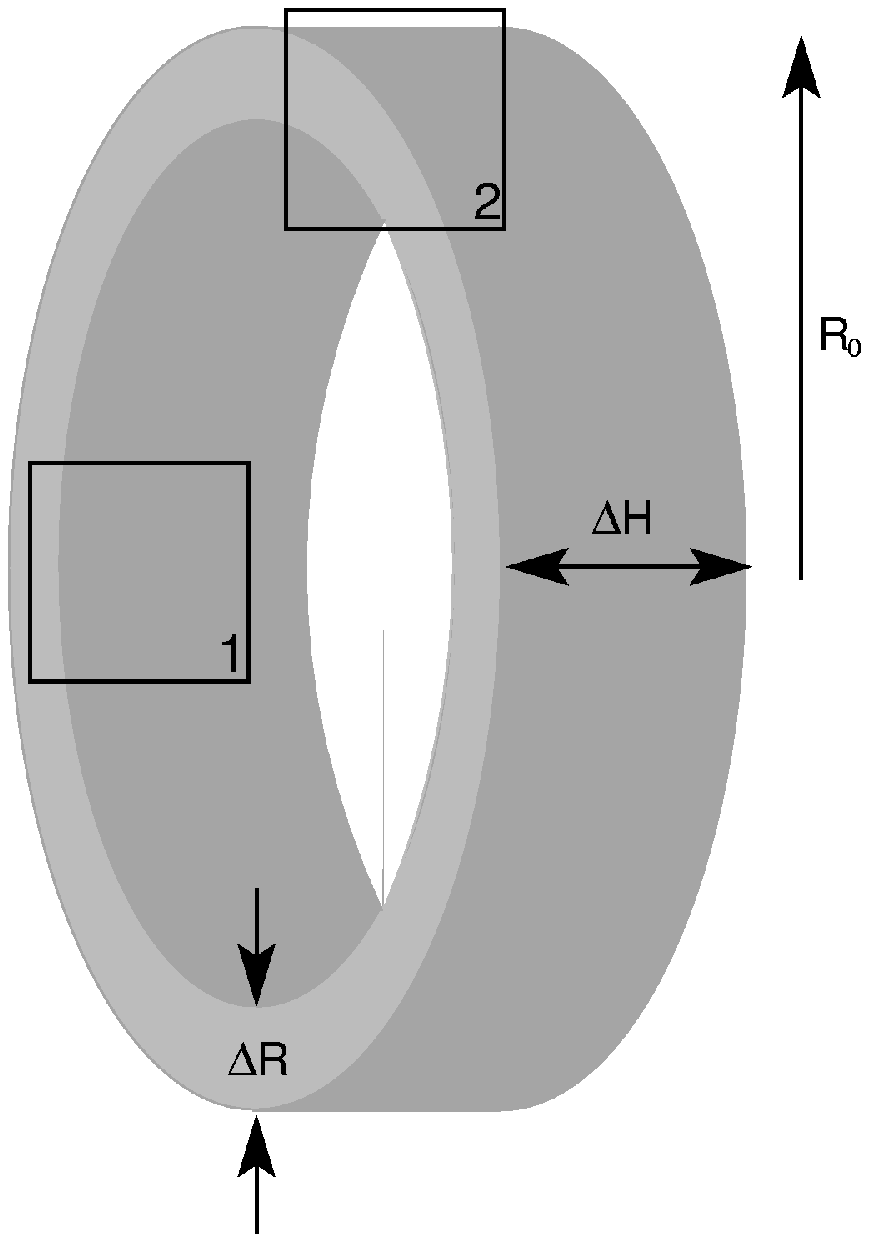}&
\includegraphics[angle=0,width=0.40\columnwidth]{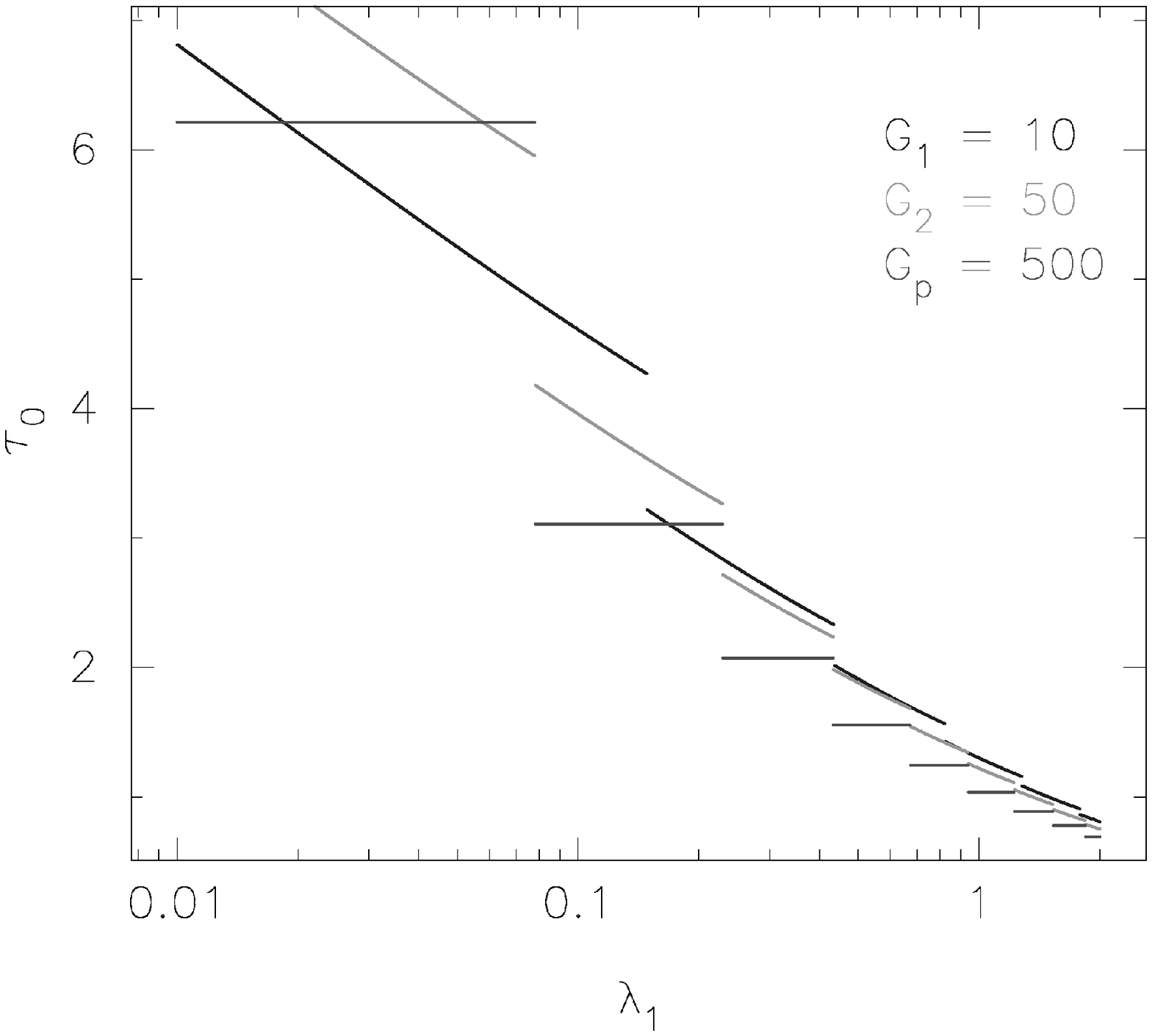}\\
\end{array}$
\end{center}
\caption{{\bf Left: }Diagram of the model. The boxes indicate the positions of
  regions 1 and 2. The outer radius of the continuum ring is not visible in
  this drawing. {\bf Right: }Solutions of equations (\ref{eq:parameters})
  shown in the $\lambda_{1} - \tau_{0}$ space. The intersection of the dark
  and light gray lines is the solution of the system.}
\label{fi:figure2}
\end{figure}

\section{Simulations}
\begin{figure}
\begin{center}
\includegraphics[angle=-90,width=0.85\columnwidth]{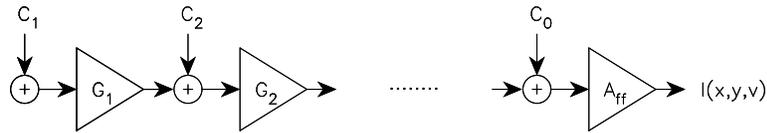}\\
\end{center}
\caption{Cascaded amplifiers model. The gain G$_{i}$ of each amplifier is
  given by $\exp\{\tau_{0}\cdot\exp((v-v_{j})^2/2\sigma^{2})\}$ Where
  $\sigma=10$ km s$^{-1}$ is the internal velocity dispersion of the clumps
  and $v_{j}$ is the projected velocity of each clump constituted by a
  turbulent and a keplerian component. C$_{i}$ are the continuum contributions 
  between the clumps and A$_{ff}$ is the free-free absorption attenuation. All
  the parameters are functions of $(x,y)$.}
\label{fi:cascade}
\end{figure}

\begin{figure}
\begin{center}
\includegraphics[angle=-90,width=0.85\columnwidth]{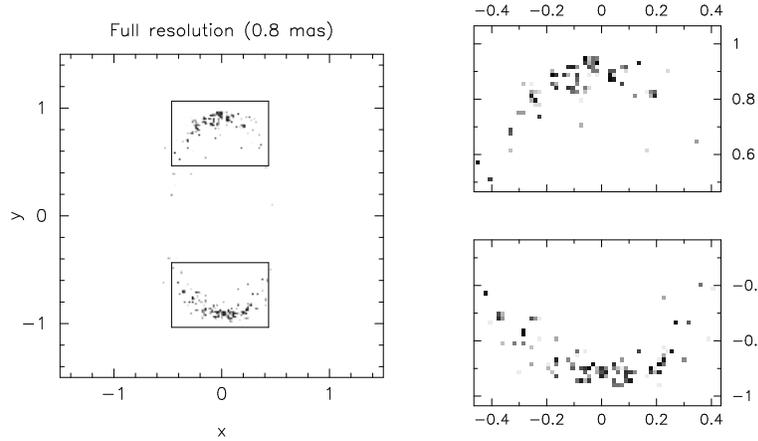}\\
\end{center}
\caption{Integrated OH intensity $I(x,y)$. The $XYV$ cube was averaged
  along all the spectral channels. The boxes indicate the northern and
  southern regions which are shown enlarged in the right panel. Both regions
  present strong compact emission dominated by a few very bright spots. In
  addition, the arc shapes observed with VLBI are fairly well reproduced.
  The grayscale has been normalized to the brightest point in the south.}
\label{fi:full_res}
\end{figure}

\begin{figure}
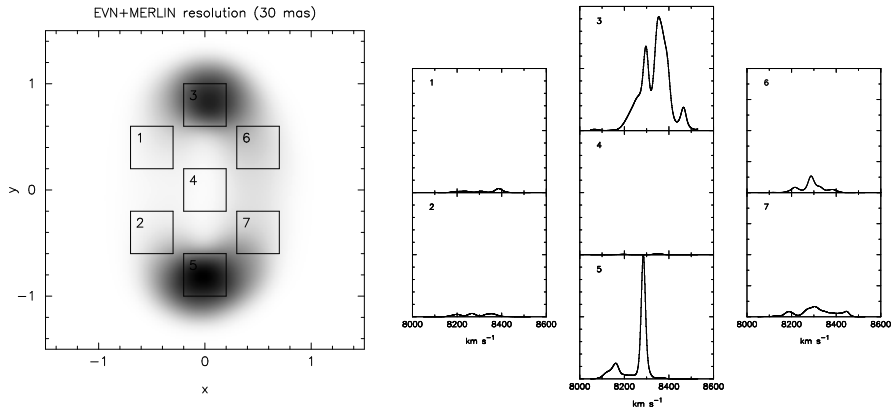

\begin{center}
$\begin{array}{c@{\hspace{0.5cm}}c}
\includegraphics[angle=-90,width=0.40\columnwidth]{map1_bw.epsi}&
\includegraphics[angle=-90,width=0.55\columnwidth]{sp.epsi}
\end{array}$
\end{center}
\caption{{\bf Left: }Integrated OH intensity convolved down to EVN+MERLIN
  resolution. The \emph{bridges} of smooth emission connecting the north and
  the south are clearly visible as well as the east-west asymmetry produced by
  the free-free absorbing cones. The grayscale has been normalized to the
  brightest point in the south. Boxes indicate the positions where spectra
  were taken. {\bf Right: }The spatial location and relative widths of the
  spectral features correspond to those observed by Pihlstr\"{o}m et
  al. 2001. The vertical scale has been normalized to the peak in the south.}
\label{fi:spectra}
\end{figure}

Using the parameters obtained in the previous section as input for Monte Carlo
simulations, artificial spectra and brightness distributions were
produced. Random generators were set up to reproduce the spatial distribution
of the individual clumps and the pseudo-turbulent velocity field. The
resulting XYZ cube was transformed into a XYV cube using the cascade model
shown in Figure \ref{fi:cascade}. The results are summarized in Figures
\ref{fi:full_res} and \ref{fi:spectra}. 

\section{Conclusions}
The proposed geometry has managed to reproduce the observed brightness
distribution and spectrum of the OH maser emission. The analytical approach in
section 2 has been useful in fine tuning the Monte Carlo simulations. However,
some aspects of the observed velocity field need to be studied. Work is still
in progress connecting the obtained results with the physics of masers and the
developing of a consistent formalism for the analysis of clumpy media.

\end{article}
\end{document}